\documentclass[twocolumn,pre,amsmath]{revtex4}

\usepackage{graphicx,setspace}
\usepackage{dcolumn}
\usepackage{bm,setspace}
\begin{document}

\title{Modeling noise induced resonance in an excitable system: An alternative approach}
\author{Md. Nurujjaman}
\email{jaman_nonlinear@yahoo.co.in}
\affiliation{ Tata Institute of Fundamental Research Centre, TIFR Centre For Applicable Mathematics, Post Bag No: 6503, Sharada Nagar, Chikkabommasandra, Bangalore 560065, India}

\begin{abstract}

Recently, it is observed [Md. Nurujjaman et al, Phy. Rev. E \textbf{80}, 015201 (R) (2009)]  that in an excitable system, one can maintain noise induced coherency in the coherence resonance by blocking the destructive effect of the noise on the system at higher noise level. This phenomenon of constant coherence resonance (CCR) cannot be explained by the existing way of simulation of the model equations of an excitable system with added noise. In this paper, we have proposed a general model which explains the noise induced resonance phenomenon  CCR as well as coherence resonance (CR) and stochastic resonance (SR).  The simulation has been carried out considering the basic mechanism of noise induced resonance phenomena: noise only perturbs the system control parameter to excite coherent oscillations, taking proper precautions so that the destructive effect of noise does not affect the system. In this approach, the CR has been obtained from the interference between the system output and noise, and the SR has been obtained by adding noise and a subthreshold signal. This also explains the observation of the frequency shift of coherent oscillations in the CCR with noise level.
 \end{abstract}

\maketitle

\section{Introduction}
Noise induced resonances: coherence resonance (CR) and stochastic resonance (SR) in a threshold or excitable systems, have been studied both numerically and experimentally in many physical, chemical, biological and electronics systems~\cite{prl:gang,prl:wiesenfeld,pre:jaman,phd:jaman,prer:jaman,prl:LinI,pop:dinklage,pre:postnov,prl:Giacomelli,prl:avila,pre:miyakawa,pre:istavan,pre:Santos1,pre:revera,revmodphys:Gammaitoni,prl:kitajo,prl:locher,pre:bascones,fnl:stock}. The regularity of the dynamical behavior of an excitable system emerges by virtue of the interplay between the autonomous nonlinear dynamics and the optimum superimposed noise that is termed as CR. Whereas, in SR, the response of the system to a weak periodic input signal  is amplified or optimized by the presence of a particular level of noise. Recently, another type of CR has been observed in plasma and electronic systems, where the coherency remains almost constant even at higher noise level~\cite{pre:jaman,phd:jaman,prer:jaman}. This CR phenomenon of constant coherency may be termed as constant coherence resonance (CCR). The main difference between CR and CCR is that, in CR autonomous dynamics is destroyed at higher noise level, whereas in CCR, system dynamics remains unaffected from the destructive effect of noise.  One of the experimentally observed feature of the CCR is that the frequency of coherent oscillations increases or decreases with increase in noise level depending on autonomous dynamics of the system. So far there is no theoretical explanation of the CCR and frequency shift.

Now the qualitative features of an excitable system can be obtained using two nonlinear autonomous differential equations which have been used for modeling the noise induced resonances~\cite{physrep:lindner}. In the existing literatures, the CR phenomenon has been modeled just by adding noise to any one of the above equations, where maximum coherency has appeared at optimum noise level~\cite{prl:gang,pre:strogatz,physrep:lindner,prl:Masoller,prl:pikovsky}. But in this approach, one cannot explain the appearance of CCR. Even explanation of SR is not very clear, though there are several experiments on this phenomenon~\cite{pre:jaman,pre:parmananda}. The main problem of adding noise directly to the equations is that the noise not only perturbs the system control parameter, but also destroys the actual dynamics. The effect of adding noise directly to the equations has been discussed in the context of the FitzHugh-Nagumo model in Section~\ref{section:result}.  In this paper, our goal is to present a general model which explains CR, SR and CCR. The model has been developed based on the excitation mechanism of the system using FitzHugh-Nagumo model, a well known paradigm for modeling noise induced resonances.

Rest of the paper has been organized as follows: we have discussed the excitation mechanism of an excitable system and effect of noise in Section~\ref{section:excitable}. Based on the physical arguments of Section~\ref{section:excitable}, we have formulated the model for the noise invoked resonances in Section~\ref{section:model}. The results and discussion of the simulation using FitzHugh-Nagumo has been presented in Section~\ref{section:result}. Finally a conclusion has been drawn in Section~\ref{section:conclusion}.

\section{Excitable dynamics and effect of noise}
\label{section:excitable}
The basic characteristic of an excitable system is that it shows a fixed point or coherent limit cycle oscillations depending on the value of the control parameter (CP) of the dynamics. The point where the system changes from oscillatory to fixed point behavior is called threshold or bifurcation point.   Now if the system shows fixed point behavior for the value of CP below the threshold, then the dynamics will be limit cycle oscillations on the other side of the threshold or vice versa, depending upon the system properties. The frequency of the autonomous oscillations in the excited state changes with control parameter. For example, frequency of the oscillations in FitzHugh-Nagumo model increases when one increases the control parameter from the threshold value. The increase in frequency of the coherent oscillations has also been observed in real experiments~\cite{pre:jaman,prer:jaman,pre:Santos1}. This has important effect on the increase in the frequency of the coherent oscillations in CCR with increase in noise level, that has been discussed at the last of next paragraph. Response of an excitable system at fixed point state to an external perturbation applied on the CP depends on the perturbation amplitude. When the amplitude is small such that the threshold is not crossed, the system remains at its fixed point state, and when it is large enough to cross the threshold,  the system returns to its fixed point deterministically, i.e., once the threshold is crossed, the system becomes almost independent of the perturbation and comes to its fixed point state traversing one limit cycle~\cite{prl:gang,prl:wiesenfeld,prl:Giacomelli,pre:strogatz,chaos:Coullet}.

For stochastic perturbation (i.e., noise), limit cycles will appear randomly due to random crossing of the threshold. The time between appearance of the consecutive limit cycles can be splitted into two characteristic times: an activation time ($t_a$) through which system resides at the fixed point state between two consecutive limit cycles, plus a refractory time ($t_r$) which is taken to traverse one limit cycle. Activation time ($t_a$) will decrease gradually with noise, as the threshold will be crossed more frequently at higher noise amplitude.  When the noise amplitude is fairly large and its changing frequency is higher than refractory frequency (which is generally the case in experiments), then the threshold will be crossed several times during one limit cycle that makes $t_a\approx 0$ and the inter-peak distances are  practically determined by the time period ($t_r$) of the limit cycle. At this stage, the dynamics is almost similar to the coherent oscillations at the excited state and is termed as coherence resonance (CR). From this stage, increase in the noise may lead to two possible effects:  firstly, it destroys the structures of the limit cycles or the actual dynamics of the system, that leads to decrease in the coherency, which has been observed in many experiments~\cite{pre:jaman,phd:jaman,prer:jaman,prl:LinI,pop:dinklage,pre:postnov,prl:Giacomelli,prl:avila,pre:miyakawa,pre:istavan,pre:Santos1,pre:revera,revmodphys:Gammaitoni,prl:kitajo,prl:locher,pre:bascones,fnl:stock}. Secondly, it perturbs the CP across the threshold so frequently that the system remains at its excited state and produces almost coherent oscillations, i.e., constant coherence resonance (CCR). CCR is not always realizable in the real experiments, as at higher noise level it is difficult to keep the noise away from its destructive effect. CCR has been observed in some recent experiments~\cite{prer:jaman,pre:jaman,phd:jaman}. Another interesting experimental observation regarding the CCR is that the frequency of  the noise invoked oscillations increases with increase in noise level in those excitable systems whose frequency of the autonomous limit cycle oscillations increases with increase in CP from the threshold~\cite{pre:jaman,phd:jaman,prer:jaman,pre:Santos1}.

Now we can summarize the effects of noise on excitable system: (a) Noise helps the system to cross the threshold by perturbing the CP; (b) it may destroy the system dynamics at higher noise level; (c) If the destructive effect of noise is blocked, the system may remain at excited state and hence may show CCR; (d) In case of CCR noise is also responsible for the effective increase in CP above the threshold. All these facts have been used to formulate a general model which has been described in the next section.

\section{Formulation of the model}
\label{section:model}

The dynamical state of an excitable system can be represented by a two dimensional state vector $X=(x,y)^T\in \Huge{R}^2$ and  the equation of motion in general form can be described by two first order differential equations  $ \frac{dX(t)}{dt}=F(X(t),a)$,  whose fixed point is given by $\frac{dX}{dt}=0=F(X)$~\cite{physrep:lindner}, where, $F$ is suitable nonlinear function and $a$ is the control parameter. Let $a=a_{th}$ be the bifurcation point (or threshold) of the system at which it goes from the fixed point to the oscillatory state. Now, the value of $a=a_0$ is so chosen that the system remains in the excitable fixed point region. Then the dynamical behaviour of the above system with Gaussian noise ($\xi$), can be written as
\begin{equation}
\label{eqn:2}
\begin{split}
 \frac{dX(t)}{dt}&=F(X(t),a_0+D\xi); ~~~for~~ D\xi+a_0>a_{th}\\
 &=0; ~~~~~~~~~~~~~~~~~~~~~~~~for~~ D\xi+a_0<a_{th}
\end{split}
\end{equation}

where, $D\xi$ represents the amplitude of the noise of standard deviation D. Here the simulation has not been carried out just adding random noise at each step of integration as was done in the previous works. The integration has been carried out based on the physics of an excitable system discussed in the previous section. Whenever CP crosses the threshold due to noise perturbation, system traverses one limit cycle. Now once the threshold is crossed, the noise remains ineffective in the dynamics until the system comes to its fixed point state, even if the CP crosses the threshold several times during this refractory phase, i.e., at the time of traversing one limit cycle~\cite{physrep:lindner}. This indicates that the noise need not be added at every step of integration during the simulation. Hence CP should be kept constant up to one limit cycle during the integration. This  is also clear from the points (a) and (c) discussed in the previous section [Section~\ref{section:excitable}]. The possible value of the CP for each limit cycle integration can be guessed from the point (d) [Section~\ref{section:excitable}]. If the noise would only lift the system at its excited state, then the frequency of the noise invoked dynamics would not change with increase in the noise level. But from the experimental results, it is clear that the frequency of the oscillations increases with increase in the noise level. So only possible reason for increase in the frequency is due to the increase in the effective value of CP. Hence, one can guess that the effective value of CP at the start of the limit cycle can be taken to be its value at fixed point state plus the noise amplitude at the time of excitation (which remains constant during one limit cycle).

Now for Eqns.~\ref{eqn:2}, whenever, the condition $D\xi+a_0>a_{th}$ is satisfied, the system will be lifted to its excited state and  $D\xi+a_0$  will remain constant during each limit cycle. During this time the system does  not feel the effect of noise, i.e., it remains independent of  noise perturbation. When it comes to its rest state, it feels again the noise perturbation and whenever, the condition $D\xi+a_0>a_{th}$ is satisfied, the system traverse one limit cycle with $a=a_0+D\xi$. By imposing this condition, noise has been blocked from its destructive effect on the system. In some cases like FitzHugh-Nagumo model, the system shows oscillatory behaviour when $a<a_{th}$. In this case the condition to be satisfied by noise is $D\xi+a_0<a_{th}$ which will be satisfied by the negative amplitudes of the noise. But the main mechanism is same for both the cases.

\section{Results and discussion}
\label{section:result}

There may be several choice of $F$ [Eqn.~\ref{eqn:2}] so that it shows an excitable dynamics~\cite{physrep:lindner}. In the present paper, the famous two dimensional Fitz Hugh-Nagumo model for excitable system has been studied, whose equations of motion are~\cite{prl:pikovsky}
\begin{equation} \label{eqn:Fitz}
\begin{split}
\frac{dx}{dt}&=\frac{1}{\epsilon}(x-\frac{x^3}{3}-y)\\
\frac{dy}{dt}&=x+a
\end{split}
\end{equation}

\begin{figure}[ht]
\centering
\includegraphics[width=8.5cm]{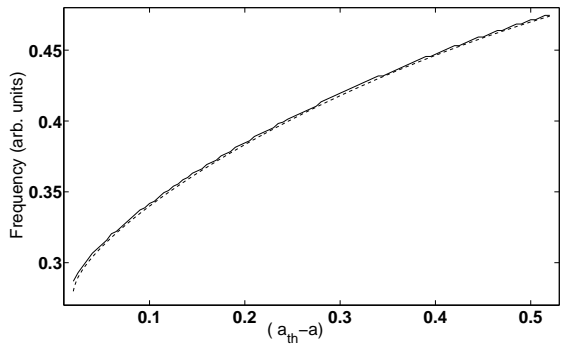}
\caption{Frequency of the limit cycle oscillations with ($a_{th}-a$) from FFT (solid line) and from inverse of the refractory time (dashed line).}
\label{fig:freq_param}
\end{figure}

where, $\epsilon=0.01$ and the control parameter `$a$' govern the dynamics. For $|a|>1$ and $|a|<1$, the system [Eqn~\ref{eqn:Fitz}] shows fixed point and limit cycle oscillations respectively. Therefore, $a_{th}=1$ is the threshold. Hence, the dynamics is fixed point and oscillatory above and below this point respectively. Frequency of the limit cycle oscillations increases between $0\leq a<1$ and then decreases between $0\geq a>-1$. In Fig~\ref{fig:freq_param}, the solid and dashed lines show increase in the frequency obtained using FFT and $t_r$ of the oscillations for parameter $1<a<0.5$ respectively and both estimations are identical. It also shows that the frequency of the autonomous dynamics increases with increase in CP  below the threshold. In the next subsection we will see that the increase in frequency of the coherent oscillation in CCR follow the trend due to noise.

\subsection{Coherence and Constant Coherence Resonance}
To study the noise invoked dynamics, `$a$' is set to $a_0=1.05$ so that Eqn~\ref{eqn:Fitz} shows fixed point behavior and is perturbed by noise ($\xi$) of standard deviation D. Whenever, $a=a_0+D\xi<1$ is satisfied, the system will traverse one limit cycle before coming to its fixed point state. The refractory time $t_r$ will depend upon the parameter value $a=a_0+D\xi$ as shown in Fig~\ref{fig:freq_param}. For low level of noise, the limit cycle appears sparsely and increases with increase in  noise level. The regularity of the appearance of these oscillations is determined by coherence parameter $NV=\sigma(T_i)/mean(T_i)$, where $\sigma$ is the standard deviations of the interpeak distances ($T_i$) of the limit cycles~\cite{prl:pikovsky}. Least $NV$ indicates high degree of coherency in the system output.

\begin{figure}[ht]
\centering
\includegraphics[width=8.5cm]{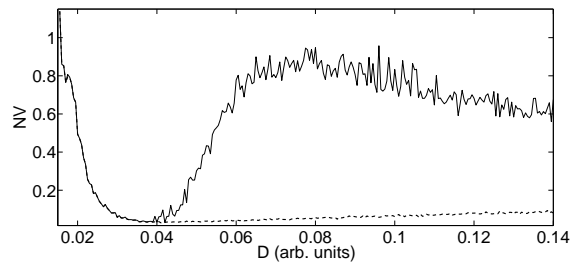}
\caption{Dashed line shows the $NV$ of the limit cycles when the destructive effect of noise is blocked. The solid line shows the NV computed after interfering the output with applied noise.}
\label{fig:nv}
\end{figure}

Fig~\ref{fig:nv} shows $NV$ vs D plot for two different cases. The dashed line [Fig~\ref{eqn:Fitz}] shows the $NV$, when the simulation was  carried out blocking the destructive effect of noise using conditions given in Section~\ref{section:model} [Eqn.~\ref{eqn:2}]. It shows that initially, $NV$ decreases rapidly with noise lavel and after reaching the minima it remains constant. As low value of $NV$ indicates coherent oscillations, the system remains in its coherent state even at higher noise level. This phenomenon may be termed as constant coherence resonance (CCR) as coherence remains unaltered at higher noise level. Though CCR has been rarely observed in experiments, it is one of the important noise induced phenomena as both CR  and SR can be obtained from the basic mechanism of CCR.

When the output of CCR was directly interfered with noise, the conventional CR has been obtained in the sense that it has unique minimum for optimum noise level. NV decreases for small level of noise  and then increases with increase in noise level as shown by solid line in Fig~\ref{fig:nv}. Minimum of the curve (solid line) represents occurrence of maximum coherence for optimum noise level. At higher level of noise NV increases due to the destructive effect of noise.  For even larger noise the output is totally governed  by noise and the jumps across the threshold are so frequent that the NV decreases again which has also been found in noise induced resonances in delayed feedback system~\cite{prl:Masoller}.

\begin{figure}[ht]
\centering
\includegraphics[width=8.5cm]{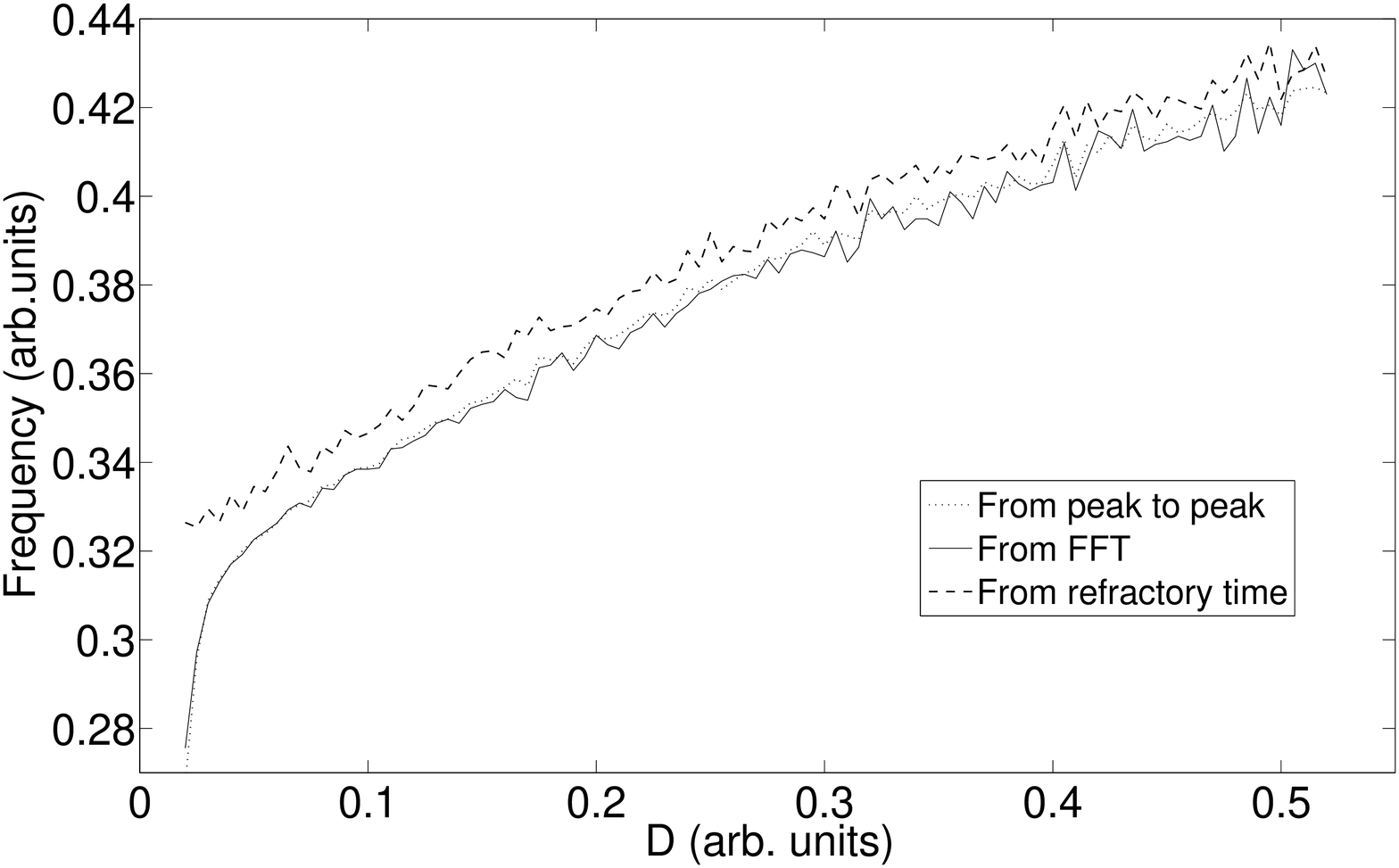}
\caption{Frequency of the limit cycles with noise: obtained from refractory time (dashed line); from interpeak distances (dot line); and from FFT (solid line).}
\label{fig:freq}
\end{figure}

In order to study the frequency change of the autonomous dynamics in CCR with noise level, we have estimated frequency using different techniques. The frequency obtained from the peak to peak distances, fast Fourier transform (FFT) and refractory time with noise has been shown in Fig~\ref{fig:freq} by using dotted, solid and dashed lines respectively. Though at small noise level, frequency obtained using above three techniques shows some difference, at higher noise level they are almost identical.  From Figs~\ref{fig:freq_param} and \ref{fig:freq}, it is clear that the trend in increase in frequency due to noise and due to increase in the CP are almost identical. This may be due to increase in effective value of CP by noise lavel.

The experimental results consistent with the above numerical results have been observed in the unijunction transistor relaxation oscillator (UJT-RO) and plasma~\cite{prer:jaman,pre:jaman,phd:jaman}. In  UJT-RO, CCR appeared due to the blocking of noise from its destructive effect by UJT itself. Detail of the experiment will be found in Ref~\cite{prer:jaman}. Whereas, in plasma, sheath was found to be a natural blocker of noise~\cite{pre:jaman,phd:jaman}.  In these experiments, the frequency of the oscillations was also observed to increase with the noise. As the frequency of the autonomous dynamics in these cases increased with increase in CP, one can assume that the effective increase in CP was due to noise and this is consistent with numerical simulation.

Earlier, in order to show CR in FitzHugh-Nagumo model, noise was added to the second Equation [of the Eqns~\ref{eqn:Fitz}] which contains the control parameter and role of noise was interpreted as an irregular modulation of the bifurcation parameter $a$ that  switches the limit cycle on and off. But  adding noise to the first equation also produces CR [Eqns~\ref{eqn:Fitz}], where switching mechanism cannot be explained~\cite{prl:pikovsky,prl:pikovsky1}. In our opinion, adding noise to any of the two equations basically destroys the actual dynamics and hence one does not get CCR. Using our approach we have got rid of this problem.

\subsection{Stochastic resonance}

In order to get stochastic resonance (SR) in the Fitz Hugh-Nagumo model, the CP has been perturbed by both a subthreshold periodic signal and noise. The main mechanism of the SR in an excitable system is that when a subthreshold signal is applied along with noise, system crosses the threshold at a place of occurrence of the peak of the subthreshold signal more frequently and hence system traverses one limit cycles at this place, i.e., the system output mimics the subthreshold signal~\cite{pre:jaman,pre:parmananda}. Here, the important point is that the structure of the limit cycles is determined by the system dynamics and the time intervals between them  are determined by interpeak distances of the applied subthreshold signal. To study SR, `$a$' has been set to $a_0=1.25$ so that it shows fixed point behavior. A subthreshold pulse of the frequency, duration and amplitude (A) 0.1, 0.05 and -0.21 respectively,  is applied along with noise $\xi$ of standard deviation D. Whenever, $a=A+D\xi+a_0<1$ is achieved, the system will traverse one limit cycle before coming to its fixed point state. The simulation is identical as for the CCR as described above except that the CP is perturbed with a subthreshold signal and noise. For low level of noise, correspondence between output and the subthreshold signal is very little, and for optimum noise level the correspondence is excellent and at higher level of noise the system response is dominated by noise.

\begin{figure}[ht]
\centering
\includegraphics[width=8.5cm]{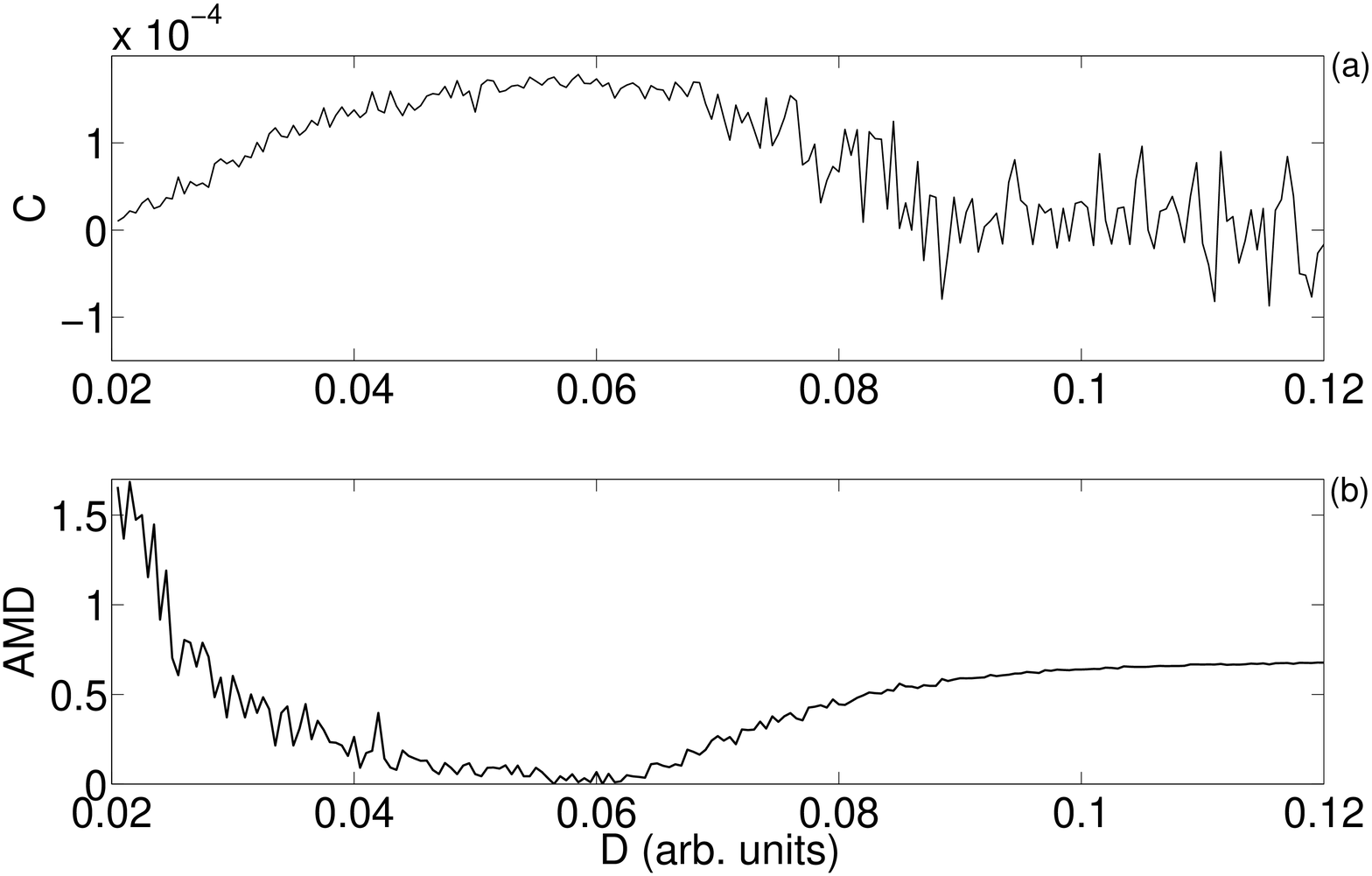}
\caption{(a) Cross correlation (C) vs D and (b) AMD vs D estimated from the time series for stochastic resonance.}
\label{fig:corr}
\end{figure}

The SR has been quantified by two stochastic parameters absolute mean difference (AMD)~\cite{prer:jaman} and correlation coefficient(C)~\cite{pre:parmananda}. AMD is defined as $AMD=abs[mean(t_p/\delta-1)]$, where, $\delta$ is the mean interpeak distance of the applied subthreshold signal. C is defined as $C=mean[(x-mean(x))(y-mean(y))]$, where, x and y are the subthreshold and output signal respectively. For SR, AMD and C should have minimum and maximum respectively. Fig~\ref{fig:corr}(a) and (b) shows C and AMD for an applied subthreshold signal of frequency 0.1. Minimum in AMD with noise and maximum in C is a typical signature of SR. The minimum corresponds to optimum noise level at which output mimics the subthreshold signal in the maximal manner. The experimental results consistent with the above numerical analysis have been found in the plasma and electro chemical system and details will be found in Ref.~\cite{pre:jaman,phd:jaman,pre:parmananda}.

\section{Conclusion}
\label{section:conclusion}
Here, we have presented an alternative modeling techniques to explain a new noise induced phenomenon CCR in an excitable system. This also explains other two noise induced resonance phenomena CR and SR. The excitation mechanism which was not clear in the earlier approach is now very much clear. Though SR and CR have been observed in many excitable system, there are a very few experiments on CCR. Possible reason for not getting CCR in the earlier experiments was that the system configurations were unable to block the destructive effect of noise. We hope that the experimental observation of CCR is possible in the earlier experiments, if the noise is allowed such that it only perturbs control parameter by choosing suitable parameters and system configurations.
\section*{acknowledgement}
I would like to acknowledge support of  Amit Apte at TIFR and some useful discussions with A.N. Sekar Iyengar. I acknowledge the use of peak detection software written by Eli Billauer ($http://billauer.co.il$). I also aknowledge the help of A. P. Choudhury to prepare the manuscript.

\end{document}